\documentclass[12pt]{article}
\usepackage{epsf}
\usepackage{a4wide}

\begin{document}
\textit{\textbf{Title:}} \\
\textbf{Comparative Molecular Dynamics and Monte Carlo Study of \\
Statistical Properties for Coarse-Grained Heteropolymers}\\[5mm]
\textit{\textbf{Authors:}} \\
Jakob Schluttig$^{1,2}$, Michael Bachmann$^1$, Wolfhard Janke$^1$\\[5mm]
\textit{\textbf{Author affiliations:}} \\
\mbox{}$^1$ Institut f\"ur Theoretische Physik and Centre for Theoretical Sciences (NTZ), 
Universit\"at Leipzig, Postfach 100\,920, D-04009 Leipzig, Germany\\
\mbox{}$^2$ Center for Modelling and Simulation in the Biosciences (BIOMS),
Ruprecht-Karls-Universit\"at Heidelberg, Im Neuenheimer Feld 267 (BIOQUANT), D-69120
Heidelberg, Germany\\[5mm]
\textit{\textbf{Corresponding author:}} \\
Dr.\ Michael Bachmann\\
Institut f\"ur Theoretische Physik, 
Universit\"at Leipzig, Postfach 100\,920, D-04009 Leipzig, Germany\\
\parbox[t]{2cm}{Phone:}\hfill\parbox[t]{16cm}{+49 341 9732443}\\
\parbox[t]{2cm}{Fax:}\hfill\parbox[t]{16cm}{+49 341 9732548}\\
\parbox[t]{2cm}{E-mail:}\hfill\parbox[t]{16cm}{Michael.Bachmann@itp.uni-leipzig.de}
\newpage
\section*{\it Abstract:}
Employing a simple hydrophobic-polar heteropolymer model, we compare thermodynamic quantities 
obtained from Andersen and Nos{\'e}-Hoover molecular dynamics as well as replica-exchange Monte Carlo
methods. We find qualitative correspondence in the results, but serious quantitative differences
using the Nos{\'e}-Hoover chain thermostat.
For analyzing the deviations, we study different parameterizations of the Nos{\'e}-Hoover chain
thermostat. Autocorrelations from molecular dynamics and Metropolis Monte Carlo runs are also
investigated. 
\\[2cm]
\textit{\textbf{Keywords:}} \\
Molecular Dynamics, Nos\'e-Hoover Thermostat, Andersen Thermostat, Replica-Exchange Monte Carlo, 
Coarse-Grained Heteropolymers
\newpage
\section{Introduction}
Understanding protein folding is one of the most complex challenges of science. The main reason is 
the characteristic property common to all proteins that the individual biological function
strongly correlates with the three-dimensional geometry of the native fold.  
Proteins are functional ingredients in all biological systems and misfolding, mutational, or
nonfunctional aggregation typically entail influential disturbances in biological networks
which frequently lead to still incurable diseases. For this reason, there is an enormously
growing interdisciplinary interest in understanding the often spontaneous
structure formation of proteins, which generally depends on the linear sequence of amino 
acids building up the macromolecule. Most of the proteins consist of hundreds of amino acids and,
therefore, thousands of atoms. The folding process is guided by usual noncovalent physical and 
chemical interactions, such as Coulomb forces due to residual charges, hydrogen bonding,
submolecular van der Waals potentials, energetic torsional barriers, and the effective 
hydrophobic force, which results as an effective interaction with the aqueous solvent
surrounding most of the non-membrane proteins. 
 
Beside the extensive bioanalytical efforts in the past few decades to resolve by X-ray and NMR 
techniques native folds of tens of thousands of bioproteins, computational simulation methods have 
become very popular tools for studying experimentally hardly manageable dynamical and thermodynamic 
properties accompanying the folding kinetics. In particular, this incorporates the central question 
of the cooperative conformational transitions the protein experiences in the formation of secondary 
structures (e.g., helices, sheets) and tertiary folds (e.g., hydrophobic-core domains).

In computational chemistry and physics, three main simulation techniques are used, which primarily 
focus on different aspects. Dynamical quantities of a process are frequently addressed by means of 
molecular dynamics (MD) methods~\cite{allen1,frenkel1}, 
where the phase-space trajectory of a system is numerically calculated 
employing a discrete, but symplectic integration scheme for the system's equations of motion. One of
the essential properties of Hamiltonian dynamics in continuum 
is that the phase-space density is constant in time.
Thus, a necessary condition for a discrete symplectic integration is that the Jacobian of the 
transformation from the system's state at time $t$ to the state at time $t+\Delta t$ is unity.
Conversely, 
Monte Carlo (MC) simulations are typically used to reveal the thermodynamic equilibrium properties of a system by 
sampling system conformations in the thermally relevant region of the phase space~\cite{bl1,berg1}. Since configurational 
updates in MC are in fact based on a Markov process, MC also possesses a kind of pseudo-dynamics. 
The third frequently used method is based on the density functional and therefore provides a tool 
for ab-initio quantum-chemical calculations of electronic structures of many-particle systems~\cite{koch1}.  
  
From a statistical physics point of view, conventional MD samples a microcanonical ensemble~\cite{gross1}, 
as the system 
energy is kept constant. However, thermal fluctuations in the physiological temperature interval are 
of essential importance for biological processes. Therefore, in more realistic simulations, the energy 
should fluctuate according to a certain, e.g., Boltzmannian distribution. In consequence, the system's
degrees of freedom have to be coupled to an environment, e.g., a heat-bath to
provide constant temperature. In our MD simulations, we used 
the stochastic Andersen thermostat~\cite{frenkel1,andersen1}
and the deterministic
Nos{\'e}-Hoover chain (NHC) thermostat~\cite{frenkel1,nose1,hoover1,martyna0}. 
It was shown for polymer systems that ensemble averages of dynamic properties are unaffected by coupling
Newtonian dynamics with Nos{\'e}-Hoover thermostats~\cite{pier1,binder1}. 
Thermostats were introduced to 
ensure that the system trajectory {\em in principle} samples 
the phase space according to the Boltzmann distribution. This means that canonical statistics should be
exactly satisfied in an infinitely long thermostated MD run. In practice, however, the question is what is
a {\em sufficiently} long run to accumulate reliable statistics. 
By definition, MD generates an extremely local ``update'' within a single time step,
whereas MC sweeps provide larger jumps in the phase space. Therefore, it is to
be expected that correlations decay much slower in MD, compared to MC.

In this paper, we aim at a comparative study of 
thermostated MD~\cite{frenkel1,andersen1,nose1,hoover1,martyna0} 
and replica-exchange MC~\cite{pt1}.
To this end, we consider a simplified coarse-grained bead-spring heteropolymer
model and concentrate on thermodynamic properties.
The paper is organized as follows. In Sec.~\ref{sec:mod}, the AB model, modified by flexible
virtual bonds, is revisited. Furthermore, we summarize the essentials of the Andersen and the NHC thermostat
used in our constant-temperature MD simulations, and the replica-exchange generalized-ensemble MC method
which is used for generating reference data. The results obtained with these methods are discussed
and the deviations are analysed in Sec.~\ref{sec:tdyn}. The paper is concluded by a summary in 
Sec.~\ref{sec:sum}.  
\section{Model and methods}
\label{sec:mod}
In this section, we introduce the model employed in our comparative study and summarize the details of the 
methods used.
\subsection{Hydrophobic-polar heteropolymer with flexible covalent bonds}
\label{sec:model}
Our study is based on a simple coarse-grained heteropolymer model, which is strongly related to the
AB model~\cite{still1}, where only hydrophobic ($A$) and hydrophilic ($B$) monomers are distinguished.  
In the original AB model, covalent bonds between adjacent monomers have a fixed length. This corresponds 
to the known stiffness of these virtual bonds~\cite{rem1}. Here, we relax this constraint through replacing 
the stiff bond by a harmonic spring. The reason is of rather technical nature as it drastically simplifies
the MD implementation using Cartesian coordinates. MD simulations of polymers with stiff bonds are 
in principle possible, e.g., by using the standard {\sc Shake}~\cite{shake1} or 
{\sc Rattle}~\cite{rattle1} algorithms.

Denoting the set of coordinates for the $N$ monomers by ${\bf R}=\{{\bf r}_1,\ldots,{\bf r}_N\}$,
we define our bead-spring variant of the AB model with $\varepsilon_0$ being an overall energy
scale as
\begin{equation}
\label{eq:model}
V({\bf R})=\varepsilon_0[v_{\rm bend}({\bf R})+v_{\rm LJ}({\bf R})+v_{\rm harm}({\bf R})],
\end{equation}
where 
\begin{equation}
\label{eq:bend}
v_{\rm bend}=\frac{1}{4}\sum\limits_{k=1}^{N-2}(1-\cos \vartheta_k)
\end{equation}
is the bending energy and the sum runs over the $(N-2)$ bending angles of successive
bond vectors. The monomer-type dependent intramolecular potential between nonbonded
monomers $i$ and $j$ with distance $r_{ij}=|{\bf r}_i-{\bf r}_j|$ 
is of Lennard-Jones form:
\begin{equation}
\label{eq:lj}
v_{\rm LJ}=4\sum\limits_{i=1}^{N-2}\sum\limits_{j=i+2}^N\left(\frac{1}{r_{ij}^{12}}
-\frac{C(\sigma_i,\sigma_j)}{r_{ij}^6} \right).
\end{equation}
The monomer-type dependence of this contribution is expressed by the parameter $C$:
\begin{equation}
\label{eq:stillc}
C(\sigma_i,\sigma_j)=\left\{\begin{array}{cl}
+1, & \hspace{7mm} \sigma_i,\sigma_j=A,\\
+1/2, & \hspace{7mm} \sigma_i,\sigma_j=B,\\
-1/2,  & \hspace{7mm} \sigma_i\neq \sigma_j.\\
\end{array} \right.
\end{equation}
The third term in Eq.~(\ref{eq:model}) is the harmonic-spring extension of the AB model and
reads:
\begin{equation}
\label{eq:harm}
v_{\rm harm}=\alpha\sum\limits_{l=1}^{N-1}\left(r_{l\, l+1}-b_0 \right)^2.
\end{equation}
The sum is taken over the $N-1$ bonds and therefore the spring energy is
related to the square deviation of the bond length from the minimum-potential distance 
$b_0$, which sets the characteristic length scale. The parameter $\alpha$ controls the bond strength and in
the strong-coupling limit $\alpha\to\infty$, the fixed-bond behavior of the original
AB model is approached. 

The kinetic energy is $E_{\rm kin}({\bf P})={\bf P}^T{\bf P}/2m$, where 
${\bf P}=\{{\bf p}_1,\ldots,{\bf p}_N\}$ is the set of the $N$ monomer momentum vectors.
Independent of their type, all monomers shall have the same mass $m$ of the order of the
average mass of an amino acid.
The Hamiltonian 
\begin{equation}
\label{eq:ham}
{\cal H}({\bf P},{\bf R})=E_{\rm kin}({\bf P})+V({\bf R})
\end{equation}
is equivalent to the total energy of the system and constant in time, 
${\cal H}({\bf P},{\bf R})\equiv E={\rm const.}$ 
Throughout the heteropolymer simulations, natural units are employed, in which
$\varepsilon_0=b_0=m=k_B=1$.

In this comparative study, we concentrate ourselves on the exemplified AB heteropolymer sequence 
S1: $BA_2BA_4BABA_2BA_5B$ with 20 monomers [14 being hydrophobic ($A$) and 6 polar ($B$)]. The thermodynamic
properties of the fixed-bond heteropolymer with this sequence have been analysed in detail 
in Ref.~\cite{baj1}. The AB model with fixed bond lengths has also proven quite useful in 
a systematic characterisation of heteropolymer folding channels known from realistic proteins~\cite{ssbj1}.
\subsection{Molecular dynamics with Andersen and Nos{\'e}-Hoover thermostat}
\label{sec:md}
Conventional molecular dynamics is governed by Newton's equations of motion, 
which read in Hamiltonian form for the $i$th particle
\begin{eqnarray}
\label{eq:hamr}
\dot{\bf r}_i&=&\nabla_{{\bf p}_i} {\cal H}({\bf P},{\bf R}) =\frac{{\bf p}_i}{m} \ ,\\
\label{eq:hamp}
\dot{\bf p}_i&=&-\nabla_{{\bf r}_i} {\cal H}({\bf P},{\bf R})=-\nabla_{{\bf r}_i}V({\bf R}).
\end{eqnarray}
System trajectories lie on a constant-energy surface, $E={\rm const.}$
In order to conserve the energy in molecular dynamics simulations with discretized
time steps, symplectic integrators are required, which provide the stability of the 
phase-space trajectory under time reversal.

In this standard form of molecular dynamics, the states of the system 
form a microcanonical ensemble in an energy
shell $E-\Delta E < E < E+\Delta E$ with $\Delta E/E\ll 1$.
The temperature does not serve as an external control parameter and is defined as 
$T=(\partial S(E)/\partial E)^{-1}$ via the microcanonical entropy $S(E)=k_B\ln \int^E_{E_{\rm min}}dE'g(E')$,
where $g(E)$ is the density of states with energy $E$~\cite{jbj1}.

In many applications, however, it is desirable to adjust the system temperature $T$
by a heat bath. Consequently, the total energies of system states follow
the canonical Boltzmann distribution $p_{\rm can}(E)\sim g(E)\exp(-E/k_BT)$. 
Typically, the folding temperature of a peptide determines the characteristic 
energy scale. 
There are mainly
two classes of approaches to introduce thermostats into Hamilton's equation of motion:
by stochastic collision forces or via virtual deterministic extensions of the phase space.
In our MD simulations, we have used the stochastic Andersen thermostat~\cite{andersen1} and the 
deterministic Nos{\'e}-Hoover chain~\cite{martyna0}. 

Using the Andersen thermostat,
a monomer experiences a random collision with a fictitious heat-bath particle after each time step $\delta t$
with probability $p_{\rm coll}=\nu\, \delta t$, where $\nu$ is the collision frequency. Therefore,
for uncorrelated random forces, the probability for a collision at time $t$ is Poissonian, 
$p_{\rm coll}(\nu,t)=\nu\exp(-\nu t)$. In case of a collision, each
component of the new momentum $p=p_{i}^{(j)}$, $j=1,2,3$, of the selected monomer
is drawn from the canonical Maxwell-Boltzmann distribution
$p_{\rm can}^{\rm kin}(p)=\exp(-p^2/2mk_BT)/\sqrt{2\pi m k_B T}$,
whose width is determined by the
temperature $T$. In Andersen dynamics, each monomer 
behaves like a Brownian particle under the influence of the external field induced by the other 
nonbonded monomers and the springs. After infinitely long time $t$, the phase-space trajectory, 
consisting of a set of nonconnected deterministic fragments, will have covered the complete
accessible phase space, which is sampled according to the Boltzmann distribution 
$p_{\rm can}(E)$. 

Nos{\'e}-Hoover dynamics is more complex, as the phase space is extended by 
$2M$ additional degrees of freedom, $\xi_k$ and $p_{\xi_k}$, $k=1,\ldots,M$.
These dynamical variables effectively represent the coupling of the system to the heat bath.
The idea is that the high-dimensional phase space provides the particles with more 
flexibility to leave the $E={\rm const.}$ trajectory in the ``true'' $({\bf P},{\bf R})$ 
phase space in a completely deterministic ``extended'' dynamics. The Nos{\'e}-Hoover energy 
\begin{equation}
\label{eq:enhc}
H_{\rm NHC}={\cal H}({\bf P},{\bf R})+\sum\limits_{k=1}^M\frac{p_{\xi_k}^2}{2Q_k}
+3Nk_BT\xi_1+k_BT\sum\limits_{k=2}^M\xi_k, 
\end{equation}
which is conserved in the multi-dimensional phase space, $H_{\rm NHC}={\rm const.}$, is defined
in such a way that by integrating out the fluctuations of the additional degrees of freedom,
the $({\bf P},{\bf R})$ states are distributed according to the canonical
ensemble.
However, the extended system is not Hamiltonian anymore and the derivation of the 
equations of motion requires some care.
Extending the Hamiltonian equations of motion~(\ref{eq:hamr}) 
and~(\ref{eq:hamp}), the Nos{\'e}-Hoover equations of motion read~\cite{frenkel1,nose1}:
\begin{eqnarray}
\label{eq:nhcr}
\dot{\bf r}_i&=&\frac{{\bf p}_i}{m}\ ,\\
\label{eq:nhcp}
\dot{\bf p}_i&=&-\nabla_{{\bf r}_i}V({\bf R})-\frac{p_{\xi_1}}{Q_1}{\bf p}_i\ ,\\
\label{eq:nhcxi}
\dot\xi_k&=&\frac{p_{\xi_k}}{Q_k}\ ,\\
\label{eq:nhcpxi1}
\dot p_{\xi_1}&=&\left(\sum\limits_{i=1}^N\frac{{\bf p}_i^2}{m}-3Nk_BT\right)-\frac{p_{\xi_2}}{Q_2}p_{\xi_1}\ ,\\
\label{eq:nhcpxik}
\dot p_{\xi_k}&=&\frac{p_{\xi_{k-1}}^2}{Q_{k-1}}-k_BT-\frac{p_{\xi_{k+1}}}{Q_{k+1}}p_{\xi_k}(1-\delta_{k M})\ .
\end{eqnarray}
The numerical values of the virtual masses $Q_k$ of the coupling variables influence the dynamics but,
in principle, not the statistical averages.
For systems, where the total energy is the only conserved quantity in the dynamics, the choice 
of a single coupling coordinate $M=1$ is sufficient. In order to destroy additional symmetries, 
however, $M>1$ couplings are required and their Nos{\'e}-Hoover equations of motion form the
linear Nos{\'e}-Hoover chain~\cite{frenkel1,martyna0}. A prominent
exceptional example is the one-dimensional harmonic oscillator, where two
coupling degrees of freedom are necessary. 

The numerical integration of the equations of motion in our simulations 
with Andersen and Nos\'e-Hoover thermostat was performed with the standard 
St\"ormer-Verlet algorithm~\cite{frenkel1,verlet1}. 
\subsubsection{Implementation details of the NHC thermostat}
In Nos\'e-Hoover dynamics, the temporal propagation for a time step $\delta t$ 
of the system and the heat-bath coupling 
degrees of freedom is governed
by the time evolution operator $U_{\rm NHC}(\delta t)$, which can be decomposed into the Trotter
factorized form
\begin{eqnarray} 
\label{eq:nhctrotter}
U_{\rm NHC}(\delta t)&=&e^{i{\cal L}_C\delta t/2}\,e^{i{\cal L}_R\delta t/2}\nonumber \\
&&\hspace*{-10mm}\times e^{i{\cal L}_P\delta t}\,
e^{i{\cal L}_R\delta t/2}\,e^{i{\cal L}_C\delta t/2}+{\cal O}(\delta t^3).
\end{eqnarray}
In this expression, ${\cal L}_P$, ${\cal L}_R$, and ${\cal L}_C$ are the Liouville operators of
the monomer momenta ${\bf P}$, the monomer coordinates ${\bf R}$, and the heat-bath coupling 
chain degrees of freedom, respectively. In higher-order integration schemes, the time step of
heat-bath coupling propagation is divided further into $n_c$ equidistant steps. Thus,
\begin{equation} 
\label{eq:nhctrotterC}
e^{i{\cal L}_C\delta t/2}=\prod\limits_{j=1}^{n_c}\prod\limits_{k=1}^m e^{i{\cal L}_Cw_k\delta t/2n_c}\ .
\end{equation}
In our NHC-MD simulations, we mainly followed the procedure described in Ref.~\cite{martyna1}.
We applied a 5th order integration scheme ($m=3$) and set $n_c=1$, i.e., the error is
of order ${\cal O}(\delta t^5)$. The Yoshida-Suzuki parameters $w_k$ are $w_{1,3}=1/(2-2^{1/3})$, 
$w_2= 1-2w_1$~\cite{yoshida1,suzuki1}. 
\subsubsection{The choice of the virtual masses for the heat-bath-coupling degrees of freedom}
\label{sec:mass}
The masses $Q_k$ ($k=1,\ldots,M$) of the virtual heat-bath ``particles'' influence the
coupling strength and, therefore, the dynamics of the correlations between the degrees of
freedom of the heat-bath and the system. Large thermal inertia $Q_k$ cause a large time scale
for the fluctuations of the heat-bath degrees of freedom. Depending on the fastest time scale
of the system, the thermostat may not be capable in balancing these fluctuations. On the
other hand, too small values of $Q_k$ {\it induce} high-frequency fluctuations into the
system, which equilibrates then very slowly. 

\begin{figure}
\centerline{\epsfxsize=17.6cm \epsfbox{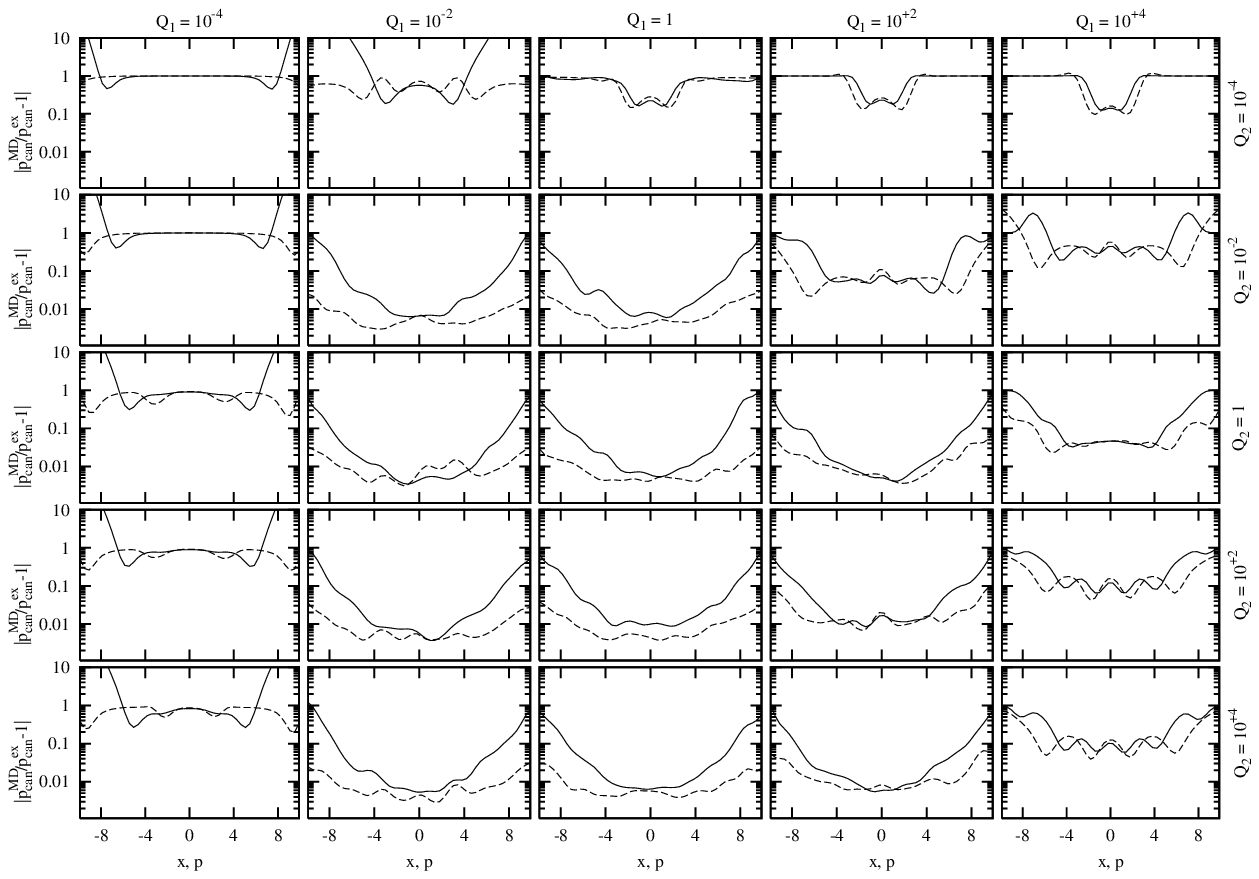}}
\caption{\label{fig:hoq}}
\end{figure}

For the one-dimensional harmonic oscillator
${\cal H}=p^2/2m+m\omega^2x^2/2$ with $m=2$ and $\omega^2=1/2$, Fig.~\ref{fig:hoq} shows 
for different choices of $Q_1$ and $Q_2$ 
the relative errors of the canonical position and momentum distributions 
as measured in NHC-MD simulations with $M=2$ Nos{\'e}-Hoover thermostats, compared with the exact 
distributions. The data were obtained by performing $10^7$ time steps of width $\delta t=0.01$ 
at $T=5$. The relative error of the measured histogram compared with the exact distribution 
is noticeably dependent on the values of $Q_1$ and $Q_2$. The biggest errors are found for 
very small and very large $Q$-values, whereas fluctuations and response seem to be balanced
much better for moderate choices of order $Q_{1,2}\sim {\cal O}(1)$. This qualitatively confirms 
the suggestion in Ref.~\cite{martyna0} to relate the $Q_k$'s to the fastest time scales
of the heat-bath (as induced by the thermal energy $\sim k_BT$) and the system ($\tau = 1/\omega$) by choosing
\begin{equation}
\label{eq:vmass}
Q_k=f_k k_BT\tau^2,
\end{equation} 
where $f_1=D\,N$ ($D$ is the spatial dimension) and 
$f_{k>1}=1$. As we can also infer from Fig.~\ref{fig:hoq},
our NHC-MD implementation works quite well for the one-dimensional harmonic oscillator with the 
properly adjusted virtual masses ($Q_1=Q_2=10$ in our units). 

For more complex systems, the identification of $\tau$
is not obvious. It can be related, e.g., to the fastest fluctuations and thus the 
largest mode in the spectrum of autocorrelation functions. The normalized 
autocorrelation function of a time-dependent quantity $s(t)$ is defined as
\begin{equation}
\label{eq:acf}
A_s(\Delta t)=\frac{\langle s(t)s(t+\Delta t) \rangle-\langle s(t)\rangle^2}%
{\langle s(t)^2\rangle-\langle s(t)\rangle^2}\ ,
\end{equation} 
where $\langle \ldots\rangle$ is the temporal average over the time series, which in
equilibrium is identical with the statistical ensemble average.

\begin{figure}
\centerline{\epsfxsize=8.8cm \epsfbox{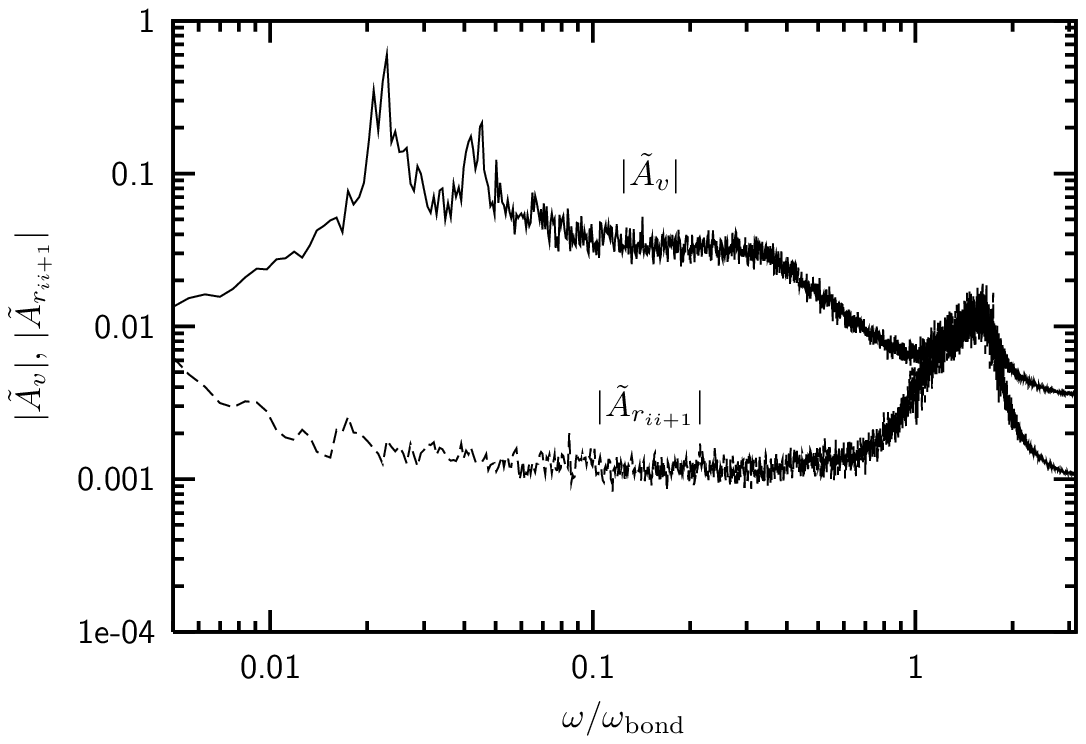}}
\caption{\label{fig:acfab}}
\end{figure}

In Fig.~\ref{fig:acfab}, the Fourier transforms, i.e., the frequency spectra,
of the velocity autocorrelation function, $\tilde{A}_v(\omega)$, and of the bond-length 
autocorrelations, $\tilde{A}_{r_{ii+1}}(\omega)$, are shown for the heteropolymer 
sequence S1. Assuming that the fluctuations of the harmonic springs~(\ref{eq:harm}) 
are of shortest time scale, the bond strength $\alpha = m\omega^2_{\rm bond}/2$ 
defines the time scale: 
\begin{equation}
\label{eq:taubond}
\tau_{\rm bond}=\omega_{\rm bond}^{-1}=\sqrt{\frac{m}{2\alpha}}.
\end{equation} 
The results for the autocorrelations obtained with an exemplified NHC-MD run at $T=1.0$
with $M=2$ Nos{\'e}-Hoover thermostats
as shown in Fig.~\ref{fig:acfab} justify this assumption. Both autocorrelation functions
have a peak close to $\omega/\omega_{\rm bond}=1$, which is the highest-frequency mode.
Therefore, we use $\tau_{\rm bond}$ for adjusting the virtual masses in Eq.~(\ref{eq:vmass})
in our NHC-MD heteropolymer simulations.
\subsection{Replica-exchange Monte Carlo method}
\label{sec:pt}
For verifying the statistical results obtained with NHC-MD of the heteropolymer model, 
a comparison with exact results is not possible. Therefore, we use a standard Markov chain 
Monte Carlo algorithm as reference method. Since the dynamics of conventional Metropolis 
sampling at fixed temperature notoriously slows down close to temperatures, where 
conformational transitions occur and also, in the dense polymer limit, a generalized-ensemble
method is more appropriate for a comprising study of thermodynamics. A simple and efficient
sampling scheme is provided by the replica-exchange or parallel tempering method~\cite{pt1}.
In this method, threads of Markov chains at different, deliberately chosen temperatures 
run in parallel and frequent trials to exchange conformations between the threads 
ensure a reasonable sampling of the accessible conformational space. In most applications,
MC methods serve to accumulate statistics by sampling regions of the configurational
space that dominate the ensemble at a given temperature. Although sometimes also employed 
for kinetic studies, the MC dynamics of the system is typically of less importance. 
For most quantities of interest, the kinetic part in the Hamiltonian~(\ref{eq:ham}) does
not influence statistical averages, i.e., the momentum fluctuations can be integrated out 
exactly. Therefore, in our MC simulations, the system experiences only coordinate
updates. In a replica-exchange step, the actual conformation ${\bf R}$ with reciprocal 
thermal energy $\beta=1/k_BT$ is tried to be exchanged with the polymer conformation ${\bf R}'$ 
being currently present in a neighbor thread running at $\beta'=1/k_BT'$. The acceptance probability 
for this exchange is simply given by
\begin{equation}
\label{eq:ptacc}
w({\bf R}\leftrightarrow {\bf R}';\beta,\beta')={\rm min}\left(1,e^{-\Delta}\right),
\end{equation}
where $\Delta = (\beta'-\beta)[V({\bf R})-V({\bf R}')]$. Hence, for $\Delta < 0$, the
exchange is always accepted. If $\Delta > 0$, the exchange is accepted with the Boltzmann-like
probability $e^{-\Delta}$. A reasonable acceptance rate can only be achieved, if
the canonical histograms of the system have sufficient overlap at the exchange temperatures
$T$ and $T'$. Therefore, the efficiency of the method strongly depends on the careful
choice of the number of threads and the associated temperatures. Conformational updates between the
exchange trials within a thread at constant temperature are accepted according to the Metropolis 
transition probability. In fixed-bond simulations conformational changes were performed using 
spherical-cap updates~\cite{baj1}. For simulations of the spring model, we used simple
Cartesian updates, where a monomer or a bond is moved. An enormous advantage of the method 
is that it can easily be parallelized as only the temperatures between the threads, running 
on individual processors, have to be communicated.
\section{Thermodynamics of S1: Comparison of results from MD and MC simulations}
\label{sec:tdyn}
In the following, we analyse the thermodynamic behavior of the flexible-bond model
for the heteropolymer sequence S1 from results obtained with Andersen MD (A-MD), 
Nos{\'e}-Hoover chain MD (NHC-MD), standard Metropolis MC (M-MC), and 
replica-exchange MC (RE-MC) simulations. The methods and their specifications in the 
simulations are listed in Table~\ref{tab:meth}.  
\subsection{Energetic and conformational fluctuations}
Energetic fluctuations are expressed by the specific heat per monomer via 
$C_V(T)=(\langle {\cal H}^2\rangle-\langle {\cal H} \rangle^2)/Nk_BT^2$. For Hamiltonian 
systems in three dimensions~(\ref{eq:ham}), this can be written as 
\begin{equation}
\label{eq:cv}
C_V(T)=C_V^{\rm kin}+\frac{1}{Nk_BT^2}\left(\langle V({\bf R})^2\rangle- 
\langle V({\bf R})\rangle^2\right),
\end{equation}
with the constant kinetic contribution $C_V^{\rm kin}=3k_B/2$. 
In our analyses, we consider only the potential energy contribution allowing for
a direct comparison with results from MC simulations.

Frequently used conformational quantities in polymer physics are the end-to-end distance
\begin{equation}
\label{eq:ree}
R_{\rm ee}({\bf R}) = |{\bf r}_N-{\bf r}_1|
\end{equation}
and the radius of gyration
\begin{equation}
\label{eq:rgyr}
R_{\rm gyr}({\bf R}) = \sqrt{\frac{1}{N}\sum\limits_{i=1}^N({\bf r}_i-{\bf r}_0)^2},\quad 
{\bf r}_0=\frac{1}{N}\sum\limits_{i=1}^N{\bf r}_i,
\end{equation}
which is a measure for the compactness of the polymer. In particular, the fluctuations 
\begin{equation}
\label{eq:rfl}
\frac{\partial}{\partial T}\langle R_{\rm ee,gyr}\rangle=\frac{1}{k_BT^2}
\left(\langle R_{\rm ee,gyr}E\rangle-\langle R_{\rm ee,gyr}\rangle\langle E\rangle\right)
\end{equation}
are of interest, as divergences or extremal points in their temperature dependence are signals for
cooperative conformational activity, i.e., they indicate conformational transitions.
It should be noted that for finite-length systems, as heteropolymers definitely are, the
fluctuations do not collapse at a certain transition temperature. Rather, different quantities
signalize activity typically at different temperatures forming a {\it transition region}~\cite{bj1}.
\subsection{Thermodynamic properties of S1}
\begin{figure}
\centerline{\epsfxsize=8.8cm \epsfbox{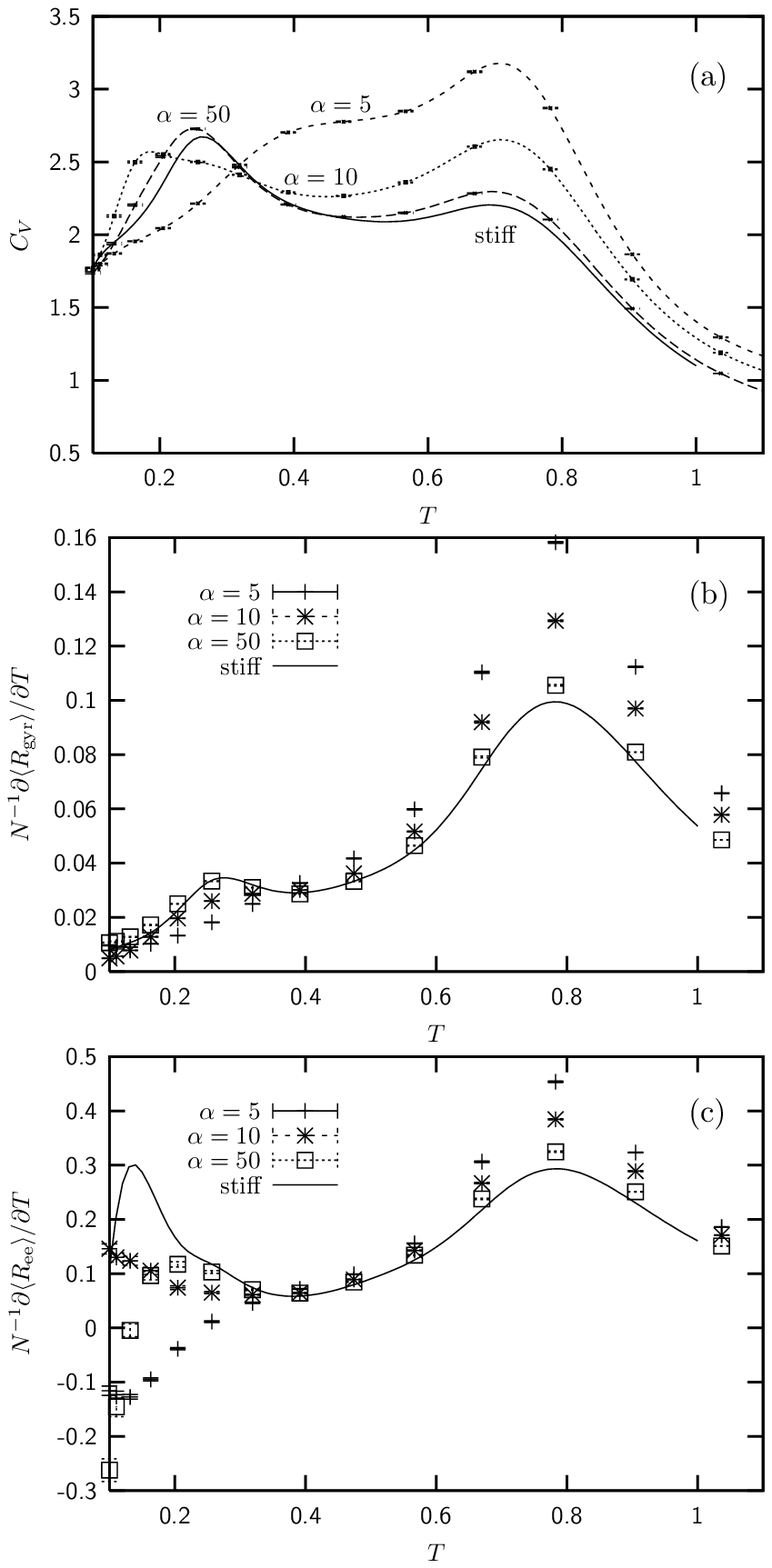}}
\caption{\label{fig:tdynflex}}
\end{figure}

As a first application, we discuss how the thermodynamic behavior of the heteropolymer
with sequence S1 depends on the flexibility of the virtual covalent bonds. 
In Figs.~\ref{fig:tdynflex}(a)--(c),
the specific heat as well as the fluctuations of gyration radius and end-to-end distance are shown
for different choices of the harmonic coupling strengths $\alpha=5,10,50$. These results were obtained
from RE-MC simulations with high precision and serve as reliable reference data. 
For large coupling strengths, these 
data reproduce former results obtained in studies for heteropolymers with stiff bonds, 
where other sophisticated MC techniques were employed~\cite{baj1,ssbj1}.
Error bars shown were calculated
using the jackknife binning method~\cite{jack1}. In all plots, there 
is a peak around $T\approx 0.8$. Although the maximum values depend on $\alpha$, the peak temperature does
not. In fact, these peaks are indications for the conformational transition between random-coil structures
and globular conformations and is, therefore, seen in all fluctuations. There is also a second region 
of activity at lower temperatures, with a comparatively weak signal in 
$\partial \langle R_{\rm gyr}\rangle/\partial T$. Actually, the heteropolymer does not experience
a further collapse, but rather an energetically favorable rearrangement of the monomers. This is an
indication for the typical heteropolymer-specific effect of the formation of a compact hydrophobic core.
The increasing strength of this signal for stiffer bonds can thus be explained with the larger barrier 
associated with the monomer rearrangement. This is not surprising as the fluctuation width of the 
springs decreases, i.e., the $N-1$ bond degrees of freedom are ``frozen''. This effect is maximal for fixed 
bonds~\cite{baj1}, where the heteropolymer possesses only $3N-(N-1)=2N+1$ degrees of freedom. It should 
be noted that, due to the different number of degrees of freedom, the specific heat even for the spring 
model with extremely stiff, but not fixed, bond lengths $\alpha\to\infty$ differs from the fixed-bond 
case by $k_B/2$ per bond. In Fig.~\ref{fig:tdynflex}, the curves for the fixed-bond case~\cite{baj1} 
are included for comparison (for the associated specific heat including the ``ficitious'' constant
offset per monomer, $(N-1)k_B/2N$, compensating the frozen-bond constraint).  

\begin{figure}
\centerline{\epsfxsize=8.8cm \epsfbox{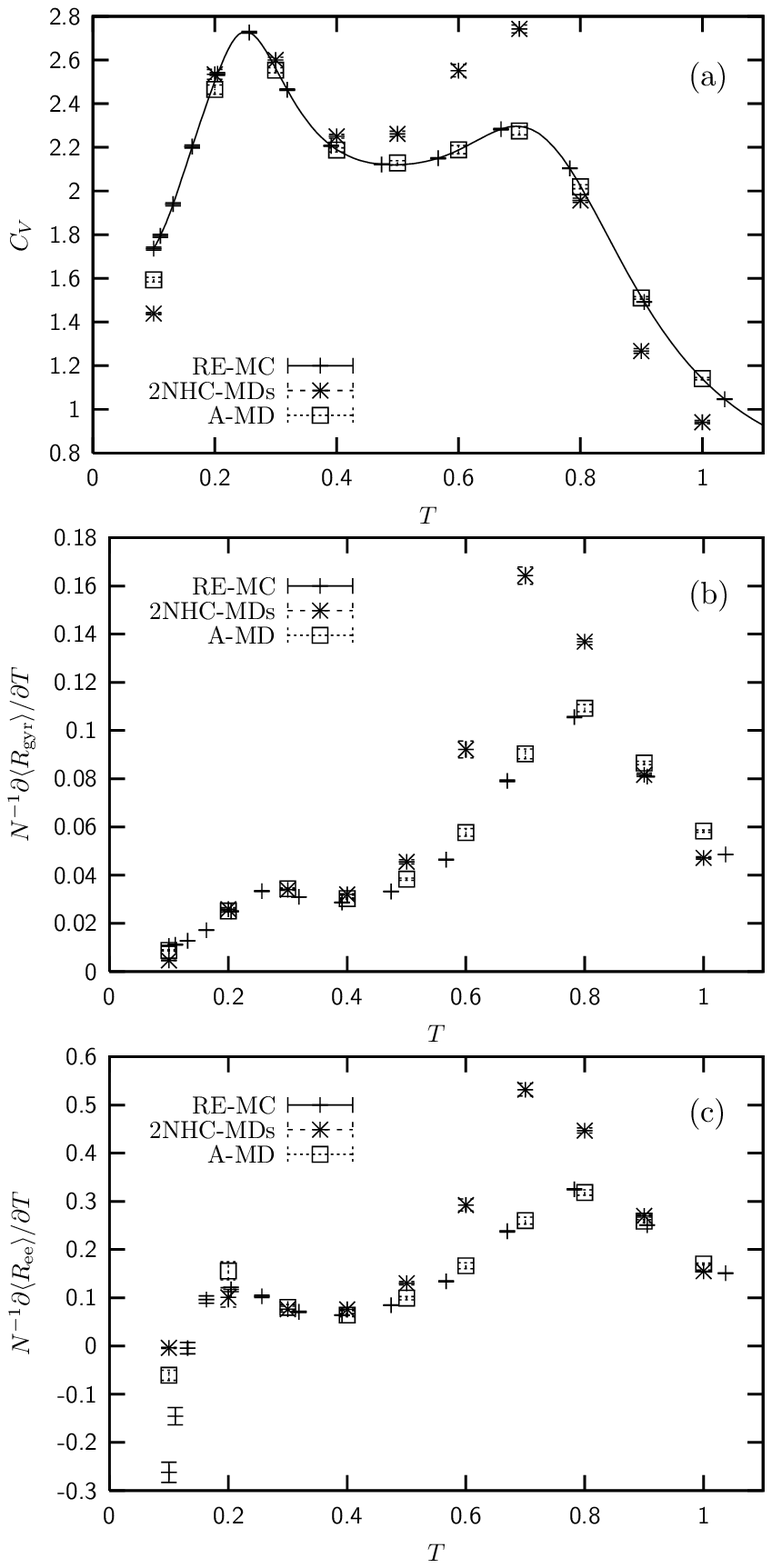}}
\caption{\label{fig:tdyncomp}}
\end{figure}

For comparing thermodynamic quantities, we performed 
RE-MC simulations and respective A-MD and 2NHC-MDs for S1 with relatively stiff bonds,
$\alpha=50$. 
The results for the specific heat and
the fluctuations of gyration radius and end-to-end distance are shown
in Figs.~\ref{fig:tdyncomp}(a)--(c), respectively. As can clearly be seen, RE-MC and A-MD
results coincide for all quantities for a wide range of temperatures. This is obviously
not the case for the 2NHC-MDs data points which deviate for temperatures $T>0.4$ 
seriously from the RE-MC results. The qualitative thermodynamic behavior, i.e., the
occurrence of conformational transitions, is still identified in all cases and the peak temperatures 
are comparable, but the quantitative agreement for the fluctuating quantities between
RE-MC and 2NHC-MDs is very poor.

\begin{figure}
\centerline{\epsfxsize=8.8cm \epsfbox{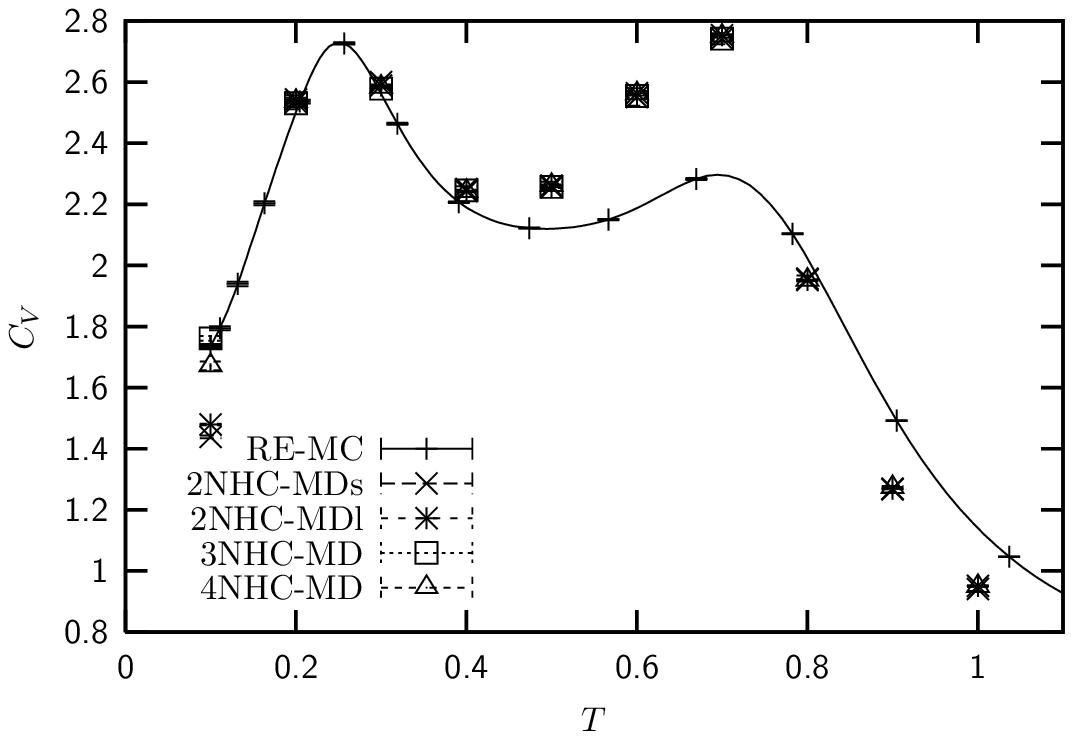}}
\caption{\label{fig:tdynnhc}}
\end{figure}

In order to see whether the deviation is due to a too
small length of the Nos{\'e}-Hoover chain, we repeated the NHC-MD simulations with
$M=3$ and $M=4$ thermostats. 
A noticeable change of the results was, however, not expected as 
no model-specific additional conserved quantities were identified.
The simplicity of the model allowed for extremely long equilibration times and run
lengths, as listed in Table~\ref{tab:meth}.
Differences between the independent MD runs, worth to be discussed in detail,
were not found. This is confirmed by the results shown in
Fig.~\ref{fig:tdynnhc}, where the output of the several tested NHC-MD variants
for the specific heat is plotted. In particular, increasing the run time of
2NHC-MDs by a factor of 20 (2NHC-MDl) does not improve the results noticeably. 

\begin{figure}
\centerline{\epsfxsize=8.8cm \epsfbox{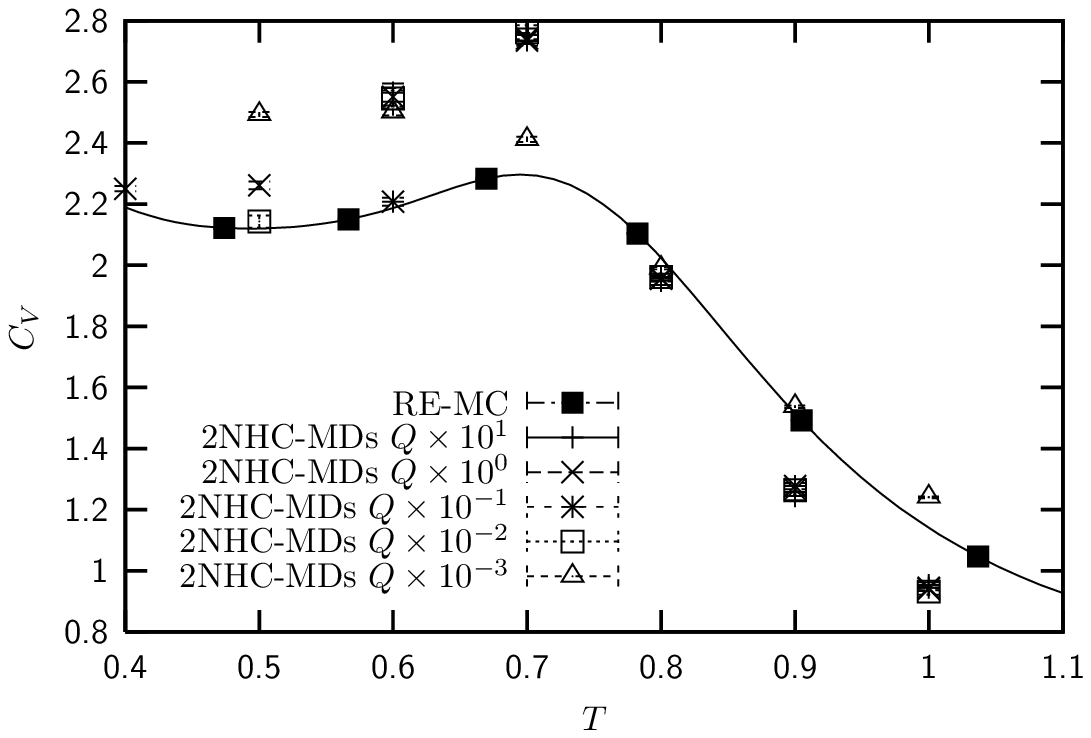}}
\caption{\label{fig:mass}}
\end{figure}

Another check we performed was to change the virtual masses. From the analysis of the 
dependence of the statistical results on the masses for the one-dimensional harmonic oscillator 
in Sec.~\ref{sec:mass}, it is clear that a careful choice of the $Q_k$ values is required to 
obtain reliable results. 
The simulations leading to the 2NHC-MDs results shown in Figs.~\ref{fig:tdyncomp}(a)--(c)
and Fig.~\ref{fig:tdynnhc} were based on the relation~(\ref{eq:vmass}) as suggested in Ref.~\cite{martyna0}. 
In order to see the influence of the virtual masses on the results, we used in additional simulations
different simple rescalings of the $Q_k$ values given by Eqs.~(\ref{eq:vmass}) and~(\ref{eq:taubond}), i.e.,
for our polymer with $\alpha=50$, $m=1$, and $N=20$, we chose as reference values
$Q_1(T)=0.6\,T$ and $Q_2(T)=0.01\,T$. 
The results for the
specific-heat estimates are shown in Fig.~\ref{fig:mass}, again compared with the replica-exchange MC
values. Actually, the results depend on the virtual masses, but simple rescaling obviously does not
solve the problem: The ``best'' choice for high temperatures, $\tilde{Q}_k=Q_k/1000$, produces wrong
results in the intermediate temperature region. It seems that the temperature dependence of the 
optimal $Q_k$'s
is nontrivial and the assumption of a linear $T$-dependence in Eq.~(\ref{eq:vmass}) is insufficient. If no data
are available for comparison, the quality of the 2NHC-MDs results cannot be appraised. This is, of course,
a substantial problem.

In further tests, it also turned out that the 2NHC-MDs results depend on the initially chosen 
conformation, i.e., the initial condition for the Nos{\'e}-Hoover dynamics was not forgotten throughout 
the run. In consequence, the thermodynamic equilibrium state space was not sampled reliably. 
This aspect is directly connected with the general MD heteropolymer folding problem: Starting from
an unfolded, denatured conformation, the native, folded state was rarely found. On the other hand,
also unfolding from the initialized native conformation is slow, but the sampling in intermediate
and high temperature regions is clearly improved. From the above results, it is not surprising
that the strongest deviations, compared with the RE-MC results, are noticed close to the collapse transition,
where an appropriate sampling of the collapsed {\em and} the random-coil phase is required.   

\begin{figure}
\centerline{\epsfxsize=8.8cm \epsfbox{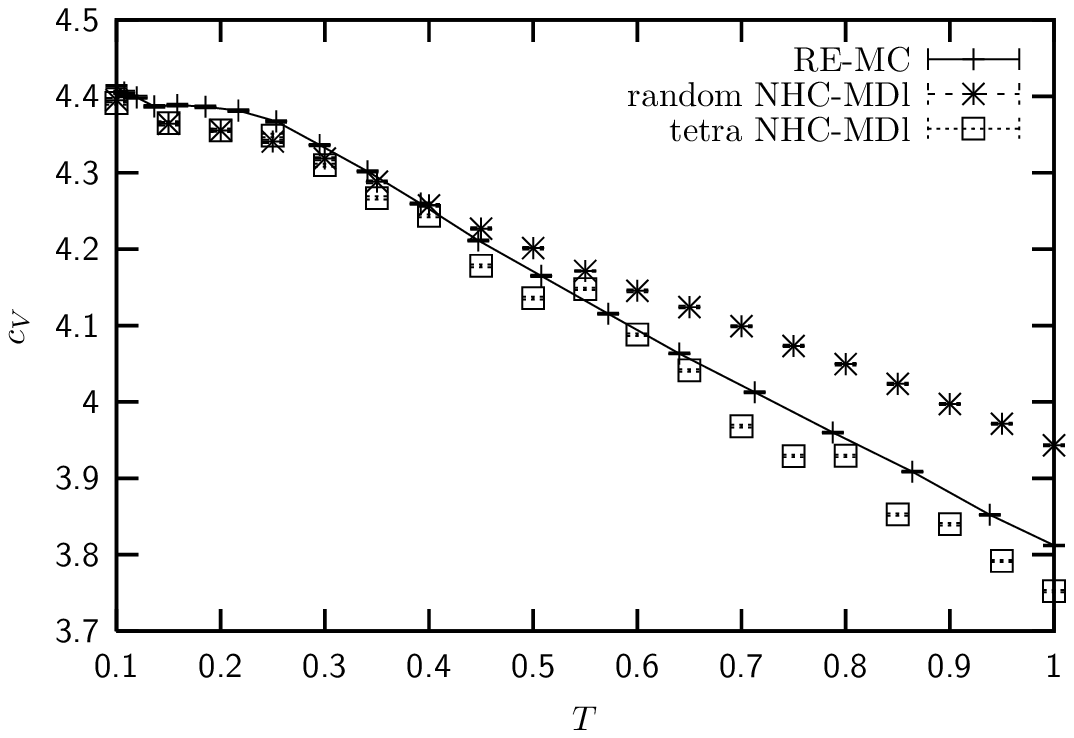}}
\caption{\label{fig:anhcv}}
\end{figure}

We have also repeated the simulations for a small polymer with 
anharmonic interactions among non-bonded monomers in order to find out whether 
systematic deviations of statistical quantities are also present in this much
simpler system.  The potential energy of this system,
\begin{equation}
\label{eq:anhA}
V({\bf R})=v_{\rm harm}({\bf R})+v_{\rm anharm}({\bf R}),
\end{equation}
consists of the harmonic bond energy $v_{\rm harm}$ as already defined in Eq.~(\ref{eq:harm})
and an anharmonic potential $v_{\rm anharm}$ for the
interaction between non-bonded monomers:
\begin{equation}
\label{eq:anhB}
v_{\rm anharm}=
\sum\limits_{l=1}^{N-1}\sum\limits_{m=l+1}^N\left[ 
\gamma_1 (r_{l\, m}-b^{\rm non}_0)^2 +\gamma_2  (r_{l\, m}-b^{\rm non}_0)^4\right].
\end{equation}
In our simulations, the model was parametrized by setting 
$\alpha=50$, $\gamma_1=10$, and $\gamma_2=1$. The
equilibrium distance between bonded and non-bonded monomers was $b_0=b^{\rm non}_0=1$.
The simulations were perfomed for a small homopolymer with $N=5$ monomers (sequence $A_5$). 
Again, statistics of parallel tempering MC
simulations was compared to results from MD simulations with NHC thermostat. The
particular parameters and run lengths were chosen according to the values
given for RE-MC and 2NHC-MDl in Table~\ref{tab:meth}. Two types of initial
conformations were used for the MD simulations. In one set of simulations, we started
from random conformations at each temperature, while in the other case
low-energy crystalline conformations were chosen, constructed from a tetrahedron with
an additional monomer mirrored at one face.  
In Fig.~\ref{fig:anhcv}, results for the specific heats of the anharmonic 5-mer, obtained with RE-MC
and differently initialized 2NHC-MDl, are shown. Again, the results of the Nos{\'e}-Hoover MD simulations 
deviate systematically from the MC output. As for the Lennard-Jones heteropolymer S1, the deviations
become stronger at high temperatures, where the NHC coupling to the heat-bath is more relevant.
\subsection{Autocorrelation time analysis for MD and MC}
For long time intervals $\Delta t$, the autocorrelation function~(\ref{eq:acf})
decays exponentially, $A_s(\Delta t)\sim \exp(-\Delta t/\tau_{\rm exp})$, where
$\tau_{\rm exp}$ is the exponential autocorrelation time. Therefore, the autocorrelation
time is a measure for the decay rate of correlations in equilibrium and depends on the dynamics of the
algorithm which is employed to generate the time series data. The statistical significance 
of a data set is connected with small temporal correlations. This means that in a time series with $L$
entries only about $n_{\rm eff}\approx L/\tau_{\rm exp}$ 
data points are uncorrelated and determine the statistical
error of the data. For this analysis, the origin of the time series, i.e., the inherent
algorithmic time scale, is irrelevant and, therefore, the autocorrelation functions of 
time series obtained with different methods can be compared. In particular, the autocorrelation
time is a good quantitative measure for the efficiency of algorithms in sampling the
relevant state space.

For two fixed temperatures, $T=0.25$ and $T=1.0$, we have compared the autocorrelation functions
and autocorrelation times 
of standard M-MC simulations, A-MD, and 2NHC-MDs for the Lennard-Jones heteropolymer S1. 
In the M-MC case, the time difference $\Delta t$
is measured in MC sweeps, where, within a single sweep, the coordinates of all monomers are 
sequentially tried to be moved randomly. The conformational changes are accepted according 
to the Metropolis criterion, with the acceptance rate adjusted at around 50\%. 
For the MD runs, $\Delta t$ is measured in units of time steps $\delta t$.

\begin{figure}
\centerline{\epsfxsize=8.8cm \epsfbox{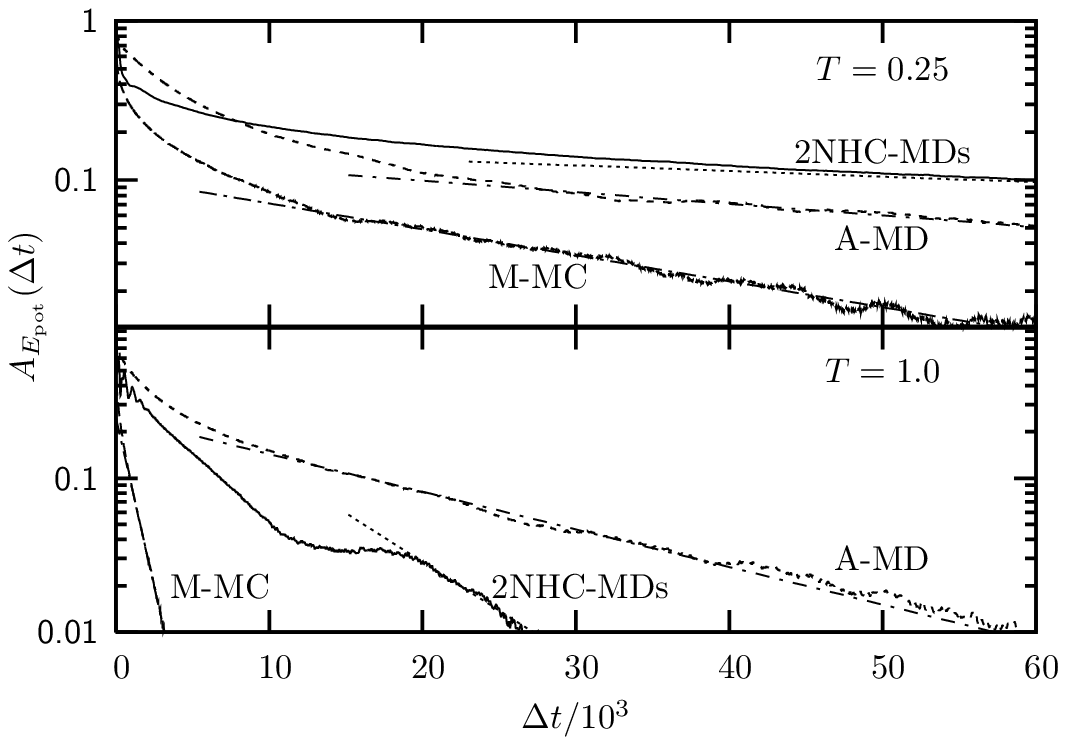}}
\caption{\label{fig:auto}}
\end{figure}

Figure~\ref{fig:auto} shows the respective autocorrelation functions of the M-MC, A-MD, 
and 2NHC-MDs simulations for $T=0.25$ and $T=1.0$. 
In Table~\ref{tab:texp}, we have listed the autocorrelation times
for the two fixed temperatures. Similarly to the MC results, the autocorrelation time in the MD runs
also decreases with temperature. The autocorrelation times differ noticeably 
for the three methods compared with each other. Not unexpectedly, autocorrelations decay
fastest for M-MC. The values of $\tau_{\rm exp}$, which are of the 
order ${\cal O}(10^3-10^5)$ MC sweeps respective MD time steps 
for the two temperatures considered, are much smaller
than the run lengths of the simulations (see Table~\ref{tab:meth}). Thus, $n_{\rm eff}\sim 10^4\ldots 10^5$
data points are uncorrelated in all MC and MD runs. For this reason, the statistical error bars for all of
our MC and MD results are very small. Thus, the partly large deviations in the
results for energetic and structural fluctuations, in particular near the conformational transitions 
(see, e.g., Fig.~\ref{fig:tdyncomp}) and for high temperatures (as in Fig.~\ref{fig:anhcv}), are 
not of statistical nature. Rather, we conclude that
the system behaves non-ergodic, i.e., not all sections of the physical phase space being thermodynamically 
relevant 
at the given temperature $T$ are covered by intersection points of the trajectory projected from 
the
extended Nos\'e-Hoover phase space. The almost constant part of the
$T=1$ 2NHC-MD curve in Fig.~\ref{fig:auto} around $\Delta t=10\,000\ldots 20\,000$ 
is also an indication that the system got stuck in a local
free-energy minimum. 
\section{Summary}
\label{sec:sum}
In this study, we have shown by explicit comparison with results from Monte Carlo
simulations that even for a minimalistic model at mesoscopic length scales 
Nos{\'e}-Hoover chain molecular dynamics simulations are not capable to reproduce the
correct thermodynamic behavior of heteropolymers. 
From our analysis, we conclude that for the polymer systems investigated
in our study, the proper stable thermodynamic equilibrium cannot be reached in
molecular dynamics simulations with Nos{\'e}-Hoover chain
thermostats and results depend on the initialization of the systems. 
In consequence, the sampling of folding and unfolding events is 
insufficient. Although the results for low temperatures are comparable with replica-exchange
Monte Carlo data, it should be noted that in the NHC-MD runs folding events were hardly
observed. Therefore, the correct formation of the hydrophobic core towards the native
fold did not happen and sampling at very low temperatures, i.e., in the hydrophobic-core
dominated region, is only reasonable, if the MD run is initialized with the native state.
For intermediate and high temperatures, we find a serious dependence of the results on the choice
of the values for the virtual masses of the heat-bath coupling degrees of freedom.  

The exemplified heteropolymer used in our study possesses only 20 monomers and is thus comparatively
small. Its folding characteristics is not particularly complex as the stiffness of the
virtual bonds has been relaxed to simplify the MD implementation. It should be noted,
however, that substituting the Nos{\'e}-Hoover thermostat against the Andersen thermostat 
with random collisions significantly improves the MD results.

In consequence, for statistical analyses of heteropolymers
in a wide range of temperatures, the applicability of canonical constant-temperature 
molecular dynamics simulations with deterministic thermostats is rather limited. 
For realistic models, the complexity of the microscopic
description at the atomic scale is known to extremely slow down NHC-MD folding simulations.
Here, however, we used a much simpler coarse-grained model and folding
events have also not been adequately recovered. 

It should be noted, however, that NHC-MD has proven to be quite successful in explaining 
dynamic processes at time scales much shorter than folding times, where, e.g., 
selected biological functions of proteins under physiological conditions can be studied.
Interesting examples, where the application of NHC-MD methods proved very useful, are
water penetration into a cell through the aquaporin membrane protein~\cite{grub1} 
and the ATP synthase, where the catalytic subunits of F1, embedded into the membrane F0 proton channel,
partially act as rotating ``molecular motor'' that promotes dehydration of ADP
and P to ATP~\cite{grub2}. Such studies require that the native folds of the proteins must 
be known as these are used as {\it input}. Substantial conformational changes of the
proteins do not occur or are limited to small segments of a few amino acids.  
\section*{Acknowledgments}
This work is partially supported by the DFG (German Science Foundation) grant  
under contract No.\ JA 483/24-1. Some simulations were performed on the 
supercomputer JUMP of the John von Neumann Institute for Computing (NIC), Forschungszentrum
J\"ulich under grant No.\ hlz11. 
\newpage
\textbf{\Large List of figure captions}\\[5mm]
\textbf{Fig.~1:} \\
Relative errors $|p_{\rm can}^{\rm MD}/p_{\rm can}^{\rm ex}-1|$ of the 
canonical position (solid line) and momentum (dashed line)
distributions $p_{\rm can}(x)$ and $p_{\rm can}(p)$ 
for a one-dimensional harmonic oscillator, estimated in NHC-MD simulations with $M=2$ for
different choices of virtual masses $Q_1$ and $Q_2$ at $T=5$.\\[3mm]
\textbf{Fig.~2:} \\
Frequency spectra of the velocity autocorrelation function $\tilde{A}_v(\omega)$ 
(upper, solid line) and of the bond-length autocorrelation $\tilde{A}_{r_{i\,i+1}}(\omega)$
(lower, dashed line) for bond strength $\alpha=50$ at $T=1$.\\[3mm]
\textbf{Fig.~3:} \\
RE-MC results for the heteropolymer S1: (a) specific heat per monomer, 
fluctuations of (b) gyration radius
and (c) end-to-end distance as functions of the temperature for different 
strengths $\alpha$ of the harmonic bonds. For comparison, also the results for a 
stiff polymer (fixed bond length) are shown [for $C_V$, the 
effect of the reduced number of degrees of freedom is artificially compensated by a constant 
offset $(N-1)/2N$]. 
Jackknife error bars are also shown, but are very small.
\newpage
\textbf{Fig.~4:} \\
Comparison of results from RE-MC, 2NHC-MDs, and A-MD simulations 
with bond strength $\alpha=50$ for the heteropolymer S1: 
(a) specific heat per monomer, 
fluctuations of (b) gyration radius
and (c) end-to-end distance as functions of the temperature.
Jackknife error bars are of symbol sizes or smaller.\\[3mm]
\textbf{Fig.~5:} \\
Results for the specific heat as obtained with the variants 2NHC-MDs,
2NHC-MDl, 3NHC-MD, and 4NHC-MD. For comparison, the RE-MC curve is also shown.\\[3mm]
\textbf{Fig.~6:} \\
Collapse transition region of the specific heat as obtained from 2NHC-MDs runs for 
different choices of the virtual 
masses $Q_k$. The reference values $Q=Q_{1,2}$ are given by Eq.~(\ref{eq:vmass}) as suggested 
in Ref.~\cite{martyna0}.\\[3mm]
\textbf{Fig.~7:} \\
Specific heats obtained from RE-MC and 2NHC-MDl runs with random and ordered start conformations
for a 5-mer with anharmonic 
interaction~(\ref{eq:anhA}),~(\ref{eq:anhB}) between non-bonded monomers.\\[3mm]
\textbf{Fig.~8:} \\
Autocorrelation functions from M-MC, A-MD, and 2NHC-MDs runs and fits
$\sim \exp(-\Delta t/\tau_{\rm exp})$ at temperatures $T=0.25$
and $T=1.0$.
\newpage
\clearpage
\newpage
\clearpage
\newpage
\clearpage
\newpage
\clearpage
\newpage
\clearpage
\newpage
\clearpage
\newpage
\begin{table*}
\caption{\label{tab:meth} Methods and specifications used in this study.
Equilibration times and run lengths are given in MC sweeps or MD steps, respectively.
}
\begin{center}
\begin{tabular}{l@{\hspace{5mm}}l@{\hspace{5mm}}c@{\hspace{5mm}}c@{\hspace{5mm}}c}\hline\hline
method & label & $M$ & equilibration & run length \\ \hline
Metropolis MC & M-MC &  & $1\times10^7$ & $1\times 10^8$ \\
Replica-Exchange MC & RE-MC & & $1\times 10^5$ & $3\times 10^8$ \\
MD with Andersen thermostat & A-MD & & $1\times 10^8$ & $3\times 10^8$ \\
MD with Nos{\'e}-Hoover chain thermostat & 2NHC-MDs & 2 & $1\times 10^8$ & $3\times 10^8$ \\
 & 2NHC-MDl & 2 & $1\times 10^9$ & $6\times 10^9$ \\
 & 3NHC-MD  & 3 & $1\times 10^9$ & $6\times 10^9$ \\
 & 4NHC-MD  & 4 & $1\times 10^9$ & $6\times 10^9$ \\ \hline \hline
\end{tabular}
\end{center}
\end{table*}
\clearpage
\newpage
\begin{table}
\caption{\label{tab:texp} Exponential autocorrelation times from M-MC, A-MD, and 2NHC-MDs runs
at $T=0.25$ and $T=1.0$. }
\begin{center}
\begin{tabular}{l@{\hspace{5mm}}l@{\hspace{5mm}}rcl}\hline\hline
& $T$ & \multicolumn{3}{c}{$\tau_{\rm exp}$} \\ \hline
M-MC     & $0.25$ &  $27$&$\times$&$10^3$\\
A-MD     & $0.25$ &  $60$&$\times$&$10^3$\\
2NHC-MDs & $0.25$ &  $124$&$\times$&$10^3$\\
M-MC     & $1.0$  &  $1$&$\times$&$10^3$\\
A-MD     & $1.0$  &  $18$&$\times$&$10^3$\\
2NHC-MDs & $1.0$  &  $7$&$\times$&$10^3$ \\ \hline \hline
\end{tabular}
\end{center}
\end{table}

\begin{thebibliography}{199}
\bibitem{rem1}
In fact, a ``covalent'' bond in the AB model between successive monomers is a virtual bond
as it actually corresponds to the distance of the C$^\alpha$ atoms of adjacent amino
acids. This virtual bond comprises three covalent bonds (C$_n^\alpha$--C'$_n$--N$_{n+1}$--C$_{n+1}^\alpha$)
and its length is nearly constant in real peptides ($\approx$ 3.8 {\AA}).
\end{thebibliography}
\end{document}